\documentclass[12pt]{article}
\usepackage{epsfig,amsfonts,amssymb}
\usepackage{hyperref}
\pdfoutput=1

\usepackage{cite}
\usepackage{cite}

\usepackage{comment}

\def\ZZZ{{\hbox{ Z\kern-1.6mm Z}}}
\def\RRR{{\hbox{ R\kern-2.4mm R}}}
\def\CCC{{\hbox{ C\kern-2.0mm C}}}
\def\zzz{{\hbox{z\kern-1mm z}}}

\newcommand{\qeq}{{\hbox{=\kern-2.3mm ? \kern.5mm }}}
\renewcommand{\qeq}{=}

\newcommand{\eps}{\epsilon}

\newcommand{\OO}{{\cal O}}

\newcommand{\be}{\begin{equation}}
\newcommand{\ee}{\end{equation}}
\newcommand{\ben}{\begin{eqnarray}\displaystyle}
\newcommand{\een}{\end{eqnarray}}

\newcommand{\refb}[1]{(\ref{#1})}
\newcommand{\p}{\partial}
\newcommand{\sectiono}[1]{\section{#1}\setcounter{equation}{0}}

\def\one{{\hbox{ 1\kern-.8mm l}}}
\def\zero{{\hbox{ 0\kern-1.5mm 0}}}

\newcommand{\bea}[1]{\begin{eqnarray}\label{#1} }
\newcommand{\eea}{\end{eqnarray}}

\newcommand{\eqref}{\refb}

\usepackage{bm}
\usepackage[table]{xcolor}

\newcommand{\non}{\nonumber}

\def\tcr{\textcolor{red}}
\def\tcb{\textcolor{blue}}

\def\tcr{\textcolor{black}}
\def\tcb{\textcolor{black}}

\def\figonea{
\def\JPicScale{0.4}
\ifx\JPicScale\undefined\def\JPicScale{1}\fi
\unitlength \JPicScale mm
\begin{picture}(135,90)(0,0)
\linethickness{0.3mm}
\multiput(45,80)(0.12,-0.12){167}{\line(0,-1){0.18}}
\linethickness{0.3mm}
\multiput(100,55)(0.12,0.12){250}{\line(1,0){0.12}}
\linethickness{0.3mm}
\multiput(100,25)(0.12,-0.12){167}{\line(1,0){0.18}}
\linethickness{0.3mm}
\multiput(40,5)(0.12,0.12){167}{\line(1,0){0.18}}
\put(120,60){\makebox(0,0)[cc]{$\bf \nearrow p_{n}$}}

\put(45,65){\makebox(0,0)[cc]{$\bf p_{1}\nwarrow$}}

\put(120,20){\makebox(0,0)[cc]{$\bf \nwarrow p'_{m}$}}

\put(42,20){\makebox(0,0)[cc]{$\bf p'_{1} \nearrow$}}

\put(80,10){\makebox(0,0)[cc]{$\cdots$}}

\put(80,70){\makebox(0,0)[cc]{$\cdots$}}

\put(80,40){\makebox(0,0)[cc]{$S$}}

\linethickness{0.3mm}
\put(105.56,40.03){\line(0,1){0.5}}
\multiput(105.55,41.03)(0.01,-0.5){1}{\line(0,-1){0.5}}
\multiput(105.53,41.53)(0.02,-0.5){1}{\line(0,-1){0.5}}
\multiput(105.5,42.04)(0.03,-0.5){1}{\line(0,-1){0.5}}
\multiput(105.46,42.54)(0.04,-0.5){1}{\line(0,-1){0.5}}
\multiput(105.41,43.04)(0.05,-0.5){1}{\line(0,-1){0.5}}
\multiput(105.35,43.54)(0.06,-0.5){1}{\line(0,-1){0.5}}
\multiput(105.28,44.03)(0.07,-0.5){1}{\line(0,-1){0.5}}
\multiput(105.2,44.53)(0.08,-0.5){1}{\line(0,-1){0.5}}
\multiput(105.11,45.02)(0.09,-0.49){1}{\line(0,-1){0.49}}
\multiput(105.01,45.52)(0.1,-0.49){1}{\line(0,-1){0.49}}
\multiput(104.9,46.01)(0.11,-0.49){1}{\line(0,-1){0.49}}
\multiput(104.78,46.5)(0.12,-0.49){1}{\line(0,-1){0.49}}
\multiput(104.65,46.98)(0.13,-0.49){1}{\line(0,-1){0.49}}
\multiput(104.51,47.47)(0.14,-0.48){1}{\line(0,-1){0.48}}
\multiput(104.37,47.95)(0.15,-0.48){1}{\line(0,-1){0.48}}
\multiput(104.21,48.42)(0.16,-0.48){1}{\line(0,-1){0.48}}
\multiput(104.04,48.9)(0.17,-0.47){1}{\line(0,-1){0.47}}
\multiput(103.87,49.37)(0.18,-0.47){1}{\line(0,-1){0.47}}
\multiput(103.68,49.84)(0.09,-0.23){2}{\line(0,-1){0.23}}
\multiput(103.49,50.3)(0.1,-0.23){2}{\line(0,-1){0.23}}
\multiput(103.28,50.76)(0.1,-0.23){2}{\line(0,-1){0.23}}
\multiput(103.07,51.21)(0.11,-0.23){2}{\line(0,-1){0.23}}
\multiput(102.85,51.66)(0.11,-0.23){2}{\line(0,-1){0.23}}
\multiput(102.62,52.11)(0.12,-0.22){2}{\line(0,-1){0.22}}
\multiput(102.38,52.55)(0.12,-0.22){2}{\line(0,-1){0.22}}
\multiput(102.13,52.99)(0.12,-0.22){2}{\line(0,-1){0.22}}
\multiput(101.87,53.42)(0.13,-0.22){2}{\line(0,-1){0.22}}
\multiput(101.61,53.85)(0.13,-0.21){2}{\line(0,-1){0.21}}
\multiput(101.33,54.27)(0.14,-0.21){2}{\line(0,-1){0.21}}
\multiput(101.05,54.69)(0.14,-0.21){2}{\line(0,-1){0.21}}
\multiput(100.76,55.1)(0.15,-0.21){2}{\line(0,-1){0.21}}
\multiput(100.46,55.5)(0.15,-0.2){2}{\line(0,-1){0.2}}
\multiput(100.15,55.9)(0.1,-0.13){3}{\line(0,-1){0.13}}
\multiput(99.84,56.29)(0.1,-0.13){3}{\line(0,-1){0.13}}
\multiput(99.52,56.68)(0.11,-0.13){3}{\line(0,-1){0.13}}
\multiput(99.19,57.06)(0.11,-0.13){3}{\line(0,-1){0.13}}
\multiput(98.85,57.43)(0.11,-0.12){3}{\line(0,-1){0.12}}
\multiput(98.5,57.79)(0.11,-0.12){3}{\line(0,-1){0.12}}
\multiput(98.15,58.15)(0.12,-0.12){3}{\line(0,-1){0.12}}
\multiput(97.79,58.5)(0.12,-0.12){3}{\line(1,0){0.12}}
\multiput(97.43,58.85)(0.12,-0.11){3}{\line(1,0){0.12}}
\multiput(97.06,59.19)(0.12,-0.11){3}{\line(1,0){0.12}}
\multiput(96.68,59.52)(0.13,-0.11){3}{\line(1,0){0.13}}
\multiput(96.29,59.84)(0.13,-0.11){3}{\line(1,0){0.13}}
\multiput(95.9,60.15)(0.13,-0.1){3}{\line(1,0){0.13}}
\multiput(95.5,60.46)(0.13,-0.1){3}{\line(1,0){0.13}}
\multiput(95.1,60.76)(0.2,-0.15){2}{\line(1,0){0.2}}
\multiput(94.69,61.05)(0.21,-0.15){2}{\line(1,0){0.21}}
\multiput(94.27,61.33)(0.21,-0.14){2}{\line(1,0){0.21}}
\multiput(93.85,61.61)(0.21,-0.14){2}{\line(1,0){0.21}}
\multiput(93.42,61.87)(0.21,-0.13){2}{\line(1,0){0.21}}
\multiput(92.99,62.13)(0.22,-0.13){2}{\line(1,0){0.22}}
\multiput(92.55,62.38)(0.22,-0.12){2}{\line(1,0){0.22}}
\multiput(92.11,62.62)(0.22,-0.12){2}{\line(1,0){0.22}}
\multiput(91.66,62.85)(0.22,-0.12){2}{\line(1,0){0.22}}
\multiput(91.21,63.07)(0.23,-0.11){2}{\line(1,0){0.23}}
\multiput(90.76,63.28)(0.23,-0.11){2}{\line(1,0){0.23}}
\multiput(90.3,63.49)(0.23,-0.1){2}{\line(1,0){0.23}}
\multiput(89.84,63.68)(0.23,-0.1){2}{\line(1,0){0.23}}
\multiput(89.37,63.87)(0.23,-0.09){2}{\line(1,0){0.23}}
\multiput(88.9,64.04)(0.47,-0.18){1}{\line(1,0){0.47}}
\multiput(88.42,64.21)(0.47,-0.17){1}{\line(1,0){0.47}}
\multiput(87.95,64.37)(0.48,-0.16){1}{\line(1,0){0.48}}
\multiput(87.47,64.51)(0.48,-0.15){1}{\line(1,0){0.48}}
\multiput(86.98,64.65)(0.48,-0.14){1}{\line(1,0){0.48}}
\multiput(86.5,64.78)(0.49,-0.13){1}{\line(1,0){0.49}}
\multiput(86.01,64.9)(0.49,-0.12){1}{\line(1,0){0.49}}
\multiput(85.52,65.01)(0.49,-0.11){1}{\line(1,0){0.49}}
\multiput(85.02,65.11)(0.49,-0.1){1}{\line(1,0){0.49}}
\multiput(84.53,65.2)(0.49,-0.09){1}{\line(1,0){0.49}}
\multiput(84.03,65.28)(0.5,-0.08){1}{\line(1,0){0.5}}
\multiput(83.54,65.35)(0.5,-0.07){1}{\line(1,0){0.5}}
\multiput(83.04,65.41)(0.5,-0.06){1}{\line(1,0){0.5}}
\multiput(82.54,65.46)(0.5,-0.05){1}{\line(1,0){0.5}}
\multiput(82.04,65.5)(0.5,-0.04){1}{\line(1,0){0.5}}
\multiput(81.53,65.53)(0.5,-0.03){1}{\line(1,0){0.5}}
\multiput(81.03,65.55)(0.5,-0.02){1}{\line(1,0){0.5}}
\multiput(80.53,65.56)(0.5,-0.01){1}{\line(1,0){0.5}}
\put(80.03,65.56){\line(1,0){0.5}}
\multiput(79.52,65.55)(0.5,0.01){1}{\line(1,0){0.5}}
\multiput(79.02,65.53)(0.5,0.02){1}{\line(1,0){0.5}}
\multiput(78.52,65.5)(0.5,0.03){1}{\line(1,0){0.5}}
\multiput(78.02,65.46)(0.5,0.04){1}{\line(1,0){0.5}}
\multiput(77.52,65.41)(0.5,0.05){1}{\line(1,0){0.5}}
\multiput(77.02,65.35)(0.5,0.06){1}{\line(1,0){0.5}}
\multiput(76.52,65.28)(0.5,0.07){1}{\line(1,0){0.5}}
\multiput(76.03,65.2)(0.5,0.08){1}{\line(1,0){0.5}}
\multiput(75.53,65.11)(0.49,0.09){1}{\line(1,0){0.49}}
\multiput(75.04,65.01)(0.49,0.1){1}{\line(1,0){0.49}}
\multiput(74.55,64.9)(0.49,0.11){1}{\line(1,0){0.49}}
\multiput(74.06,64.78)(0.49,0.12){1}{\line(1,0){0.49}}
\multiput(73.57,64.65)(0.49,0.13){1}{\line(1,0){0.49}}
\multiput(73.09,64.51)(0.48,0.14){1}{\line(1,0){0.48}}
\multiput(72.61,64.37)(0.48,0.15){1}{\line(1,0){0.48}}
\multiput(72.13,64.21)(0.48,0.16){1}{\line(1,0){0.48}}
\multiput(71.66,64.04)(0.47,0.17){1}{\line(1,0){0.47}}
\multiput(71.19,63.87)(0.47,0.18){1}{\line(1,0){0.47}}
\multiput(70.72,63.68)(0.23,0.09){2}{\line(1,0){0.23}}
\multiput(70.26,63.49)(0.23,0.1){2}{\line(1,0){0.23}}
\multiput(69.8,63.28)(0.23,0.1){2}{\line(1,0){0.23}}
\multiput(69.34,63.07)(0.23,0.11){2}{\line(1,0){0.23}}
\multiput(68.89,62.85)(0.23,0.11){2}{\line(1,0){0.23}}
\multiput(68.44,62.62)(0.22,0.12){2}{\line(1,0){0.22}}
\multiput(68,62.38)(0.22,0.12){2}{\line(1,0){0.22}}
\multiput(67.57,62.13)(0.22,0.12){2}{\line(1,0){0.22}}
\multiput(67.13,61.87)(0.22,0.13){2}{\line(1,0){0.22}}
\multiput(66.71,61.61)(0.21,0.13){2}{\line(1,0){0.21}}
\multiput(66.29,61.33)(0.21,0.14){2}{\line(1,0){0.21}}
\multiput(65.87,61.05)(0.21,0.14){2}{\line(1,0){0.21}}
\multiput(65.46,60.76)(0.21,0.15){2}{\line(1,0){0.21}}
\multiput(65.06,60.46)(0.2,0.15){2}{\line(1,0){0.2}}
\multiput(64.66,60.15)(0.13,0.1){3}{\line(1,0){0.13}}
\multiput(64.27,59.84)(0.13,0.1){3}{\line(1,0){0.13}}
\multiput(63.88,59.52)(0.13,0.11){3}{\line(1,0){0.13}}
\multiput(63.5,59.19)(0.13,0.11){3}{\line(1,0){0.13}}
\multiput(63.13,58.85)(0.12,0.11){3}{\line(1,0){0.12}}
\multiput(62.76,58.5)(0.12,0.11){3}{\line(1,0){0.12}}
\multiput(62.4,58.15)(0.12,0.12){3}{\line(1,0){0.12}}
\multiput(62.05,57.79)(0.12,0.12){3}{\line(0,1){0.12}}
\multiput(61.71,57.43)(0.11,0.12){3}{\line(0,1){0.12}}
\multiput(61.37,57.06)(0.11,0.12){3}{\line(0,1){0.12}}
\multiput(61.04,56.68)(0.11,0.13){3}{\line(0,1){0.13}}
\multiput(60.72,56.29)(0.11,0.13){3}{\line(0,1){0.13}}
\multiput(60.4,55.9)(0.1,0.13){3}{\line(0,1){0.13}}
\multiput(60.1,55.5)(0.1,0.13){3}{\line(0,1){0.13}}
\multiput(59.8,55.1)(0.15,0.2){2}{\line(0,1){0.2}}
\multiput(59.51,54.69)(0.15,0.21){2}{\line(0,1){0.21}}
\multiput(59.22,54.27)(0.14,0.21){2}{\line(0,1){0.21}}
\multiput(58.95,53.85)(0.14,0.21){2}{\line(0,1){0.21}}
\multiput(58.68,53.42)(0.13,0.21){2}{\line(0,1){0.21}}
\multiput(58.43,52.99)(0.13,0.22){2}{\line(0,1){0.22}}
\multiput(58.18,52.55)(0.12,0.22){2}{\line(0,1){0.22}}
\multiput(57.94,52.11)(0.12,0.22){2}{\line(0,1){0.22}}
\multiput(57.71,51.66)(0.12,0.22){2}{\line(0,1){0.22}}
\multiput(57.49,51.21)(0.11,0.23){2}{\line(0,1){0.23}}
\multiput(57.27,50.76)(0.11,0.23){2}{\line(0,1){0.23}}
\multiput(57.07,50.3)(0.1,0.23){2}{\line(0,1){0.23}}
\multiput(56.87,49.84)(0.1,0.23){2}{\line(0,1){0.23}}
\multiput(56.69,49.37)(0.09,0.23){2}{\line(0,1){0.23}}
\multiput(56.51,48.9)(0.18,0.47){1}{\line(0,1){0.47}}
\multiput(56.35,48.42)(0.17,0.47){1}{\line(0,1){0.47}}
\multiput(56.19,47.95)(0.16,0.48){1}{\line(0,1){0.48}}
\multiput(56.04,47.47)(0.15,0.48){1}{\line(0,1){0.48}}
\multiput(55.9,46.98)(0.14,0.48){1}{\line(0,1){0.48}}
\multiput(55.78,46.5)(0.13,0.49){1}{\line(0,1){0.49}}
\multiput(55.66,46.01)(0.12,0.49){1}{\line(0,1){0.49}}
\multiput(55.55,45.52)(0.11,0.49){1}{\line(0,1){0.49}}
\multiput(55.45,45.02)(0.1,0.49){1}{\line(0,1){0.49}}
\multiput(55.36,44.53)(0.09,0.49){1}{\line(0,1){0.49}}
\multiput(55.28,44.03)(0.08,0.5){1}{\line(0,1){0.5}}
\multiput(55.21,43.54)(0.07,0.5){1}{\line(0,1){0.5}}
\multiput(55.15,43.04)(0.06,0.5){1}{\line(0,1){0.5}}
\multiput(55.1,42.54)(0.05,0.5){1}{\line(0,1){0.5}}
\multiput(55.06,42.04)(0.04,0.5){1}{\line(0,1){0.5}}
\multiput(55.03,41.53)(0.03,0.5){1}{\line(0,1){0.5}}
\multiput(55.01,41.03)(0.02,0.5){1}{\line(0,1){0.5}}
\multiput(55,40.53)(0.01,0.5){1}{\line(0,1){0.5}}
\put(55,40.03){\line(0,1){0.5}}
\multiput(55,40.03)(0.01,-0.5){1}{\line(0,-1){0.5}}
\multiput(55.01,39.52)(0.02,-0.5){1}{\line(0,-1){0.5}}
\multiput(55.03,39.02)(0.03,-0.5){1}{\line(0,-1){0.5}}
\multiput(55.06,38.52)(0.04,-0.5){1}{\line(0,-1){0.5}}
\multiput(55.1,38.02)(0.05,-0.5){1}{\line(0,-1){0.5}}
\multiput(55.15,37.52)(0.06,-0.5){1}{\line(0,-1){0.5}}
\multiput(55.21,37.02)(0.07,-0.5){1}{\line(0,-1){0.5}}
\multiput(55.28,36.52)(0.08,-0.5){1}{\line(0,-1){0.5}}
\multiput(55.36,36.03)(0.09,-0.49){1}{\line(0,-1){0.49}}
\multiput(55.45,35.53)(0.1,-0.49){1}{\line(0,-1){0.49}}
\multiput(55.55,35.04)(0.11,-0.49){1}{\line(0,-1){0.49}}
\multiput(55.66,34.55)(0.12,-0.49){1}{\line(0,-1){0.49}}
\multiput(55.78,34.06)(0.13,-0.49){1}{\line(0,-1){0.49}}
\multiput(55.9,33.57)(0.14,-0.48){1}{\line(0,-1){0.48}}
\multiput(56.04,33.09)(0.15,-0.48){1}{\line(0,-1){0.48}}
\multiput(56.19,32.61)(0.16,-0.48){1}{\line(0,-1){0.48}}
\multiput(56.35,32.13)(0.17,-0.47){1}{\line(0,-1){0.47}}
\multiput(56.51,31.66)(0.18,-0.47){1}{\line(0,-1){0.47}}
\multiput(56.69,31.19)(0.09,-0.23){2}{\line(0,-1){0.23}}
\multiput(56.87,30.72)(0.1,-0.23){2}{\line(0,-1){0.23}}
\multiput(57.07,30.26)(0.1,-0.23){2}{\line(0,-1){0.23}}
\multiput(57.27,29.8)(0.11,-0.23){2}{\line(0,-1){0.23}}
\multiput(57.49,29.34)(0.11,-0.23){2}{\line(0,-1){0.23}}
\multiput(57.71,28.89)(0.12,-0.22){2}{\line(0,-1){0.22}}
\multiput(57.94,28.44)(0.12,-0.22){2}{\line(0,-1){0.22}}
\multiput(58.18,28)(0.12,-0.22){2}{\line(0,-1){0.22}}
\multiput(58.43,27.57)(0.13,-0.22){2}{\line(0,-1){0.22}}
\multiput(58.68,27.13)(0.13,-0.21){2}{\line(0,-1){0.21}}
\multiput(58.95,26.71)(0.14,-0.21){2}{\line(0,-1){0.21}}
\multiput(59.22,26.29)(0.14,-0.21){2}{\line(0,-1){0.21}}
\multiput(59.51,25.87)(0.15,-0.21){2}{\line(0,-1){0.21}}
\multiput(59.8,25.46)(0.15,-0.2){2}{\line(0,-1){0.2}}
\multiput(60.1,25.06)(0.1,-0.13){3}{\line(0,-1){0.13}}
\multiput(60.4,24.66)(0.1,-0.13){3}{\line(0,-1){0.13}}
\multiput(60.72,24.27)(0.11,-0.13){3}{\line(0,-1){0.13}}
\multiput(61.04,23.88)(0.11,-0.13){3}{\line(0,-1){0.13}}
\multiput(61.37,23.5)(0.11,-0.12){3}{\line(0,-1){0.12}}
\multiput(61.71,23.13)(0.11,-0.12){3}{\line(0,-1){0.12}}
\multiput(62.05,22.76)(0.12,-0.12){3}{\line(0,-1){0.12}}
\multiput(62.4,22.4)(0.12,-0.12){3}{\line(1,0){0.12}}
\multiput(62.76,22.05)(0.12,-0.11){3}{\line(1,0){0.12}}
\multiput(63.13,21.71)(0.12,-0.11){3}{\line(1,0){0.12}}
\multiput(63.5,21.37)(0.13,-0.11){3}{\line(1,0){0.13}}
\multiput(63.88,21.04)(0.13,-0.11){3}{\line(1,0){0.13}}
\multiput(64.27,20.72)(0.13,-0.1){3}{\line(1,0){0.13}}
\multiput(64.66,20.4)(0.13,-0.1){3}{\line(1,0){0.13}}
\multiput(65.06,20.1)(0.2,-0.15){2}{\line(1,0){0.2}}
\multiput(65.46,19.8)(0.21,-0.15){2}{\line(1,0){0.21}}
\multiput(65.87,19.51)(0.21,-0.14){2}{\line(1,0){0.21}}
\multiput(66.29,19.22)(0.21,-0.14){2}{\line(1,0){0.21}}
\multiput(66.71,18.95)(0.21,-0.13){2}{\line(1,0){0.21}}
\multiput(67.13,18.68)(0.22,-0.13){2}{\line(1,0){0.22}}
\multiput(67.57,18.43)(0.22,-0.12){2}{\line(1,0){0.22}}
\multiput(68,18.18)(0.22,-0.12){2}{\line(1,0){0.22}}
\multiput(68.44,17.94)(0.22,-0.12){2}{\line(1,0){0.22}}
\multiput(68.89,17.71)(0.23,-0.11){2}{\line(1,0){0.23}}
\multiput(69.34,17.49)(0.23,-0.11){2}{\line(1,0){0.23}}
\multiput(69.8,17.27)(0.23,-0.1){2}{\line(1,0){0.23}}
\multiput(70.26,17.07)(0.23,-0.1){2}{\line(1,0){0.23}}
\multiput(70.72,16.87)(0.23,-0.09){2}{\line(1,0){0.23}}
\multiput(71.19,16.69)(0.47,-0.18){1}{\line(1,0){0.47}}
\multiput(71.66,16.51)(0.47,-0.17){1}{\line(1,0){0.47}}
\multiput(72.13,16.35)(0.48,-0.16){1}{\line(1,0){0.48}}
\multiput(72.61,16.19)(0.48,-0.15){1}{\line(1,0){0.48}}
\multiput(73.09,16.04)(0.48,-0.14){1}{\line(1,0){0.48}}
\multiput(73.57,15.9)(0.49,-0.13){1}{\line(1,0){0.49}}
\multiput(74.06,15.78)(0.49,-0.12){1}{\line(1,0){0.49}}
\multiput(74.55,15.66)(0.49,-0.11){1}{\line(1,0){0.49}}
\multiput(75.04,15.55)(0.49,-0.1){1}{\line(1,0){0.49}}
\multiput(75.53,15.45)(0.49,-0.09){1}{\line(1,0){0.49}}
\multiput(76.03,15.36)(0.5,-0.08){1}{\line(1,0){0.5}}
\multiput(76.52,15.28)(0.5,-0.07){1}{\line(1,0){0.5}}
\multiput(77.02,15.21)(0.5,-0.06){1}{\line(1,0){0.5}}
\multiput(77.52,15.15)(0.5,-0.05){1}{\line(1,0){0.5}}
\multiput(78.02,15.1)(0.5,-0.04){1}{\line(1,0){0.5}}
\multiput(78.52,15.06)(0.5,-0.03){1}{\line(1,0){0.5}}
\multiput(79.02,15.03)(0.5,-0.02){1}{\line(1,0){0.5}}
\multiput(79.52,15.01)(0.5,-0.01){1}{\line(1,0){0.5}}
\put(80.03,15){\line(1,0){0.5}}
\multiput(80.53,15)(0.5,0.01){1}{\line(1,0){0.5}}
\multiput(81.03,15.01)(0.5,0.02){1}{\line(1,0){0.5}}
\multiput(81.53,15.03)(0.5,0.03){1}{\line(1,0){0.5}}
\multiput(82.04,15.06)(0.5,0.04){1}{\line(1,0){0.5}}
\multiput(82.54,15.1)(0.5,0.05){1}{\line(1,0){0.5}}
\multiput(83.04,15.15)(0.5,0.06){1}{\line(1,0){0.5}}
\multiput(83.54,15.21)(0.5,0.07){1}{\line(1,0){0.5}}
\multiput(84.03,15.28)(0.5,0.08){1}{\line(1,0){0.5}}
\multiput(84.53,15.36)(0.49,0.09){1}{\line(1,0){0.49}}
\multiput(85.02,15.45)(0.49,0.1){1}{\line(1,0){0.49}}
\multiput(85.52,15.55)(0.49,0.11){1}{\line(1,0){0.49}}
\multiput(86.01,15.66)(0.49,0.12){1}{\line(1,0){0.49}}
\multiput(86.5,15.78)(0.49,0.13){1}{\line(1,0){0.49}}
\multiput(86.98,15.9)(0.48,0.14){1}{\line(1,0){0.48}}
\multiput(87.47,16.04)(0.48,0.15){1}{\line(1,0){0.48}}
\multiput(87.95,16.19)(0.48,0.16){1}{\line(1,0){0.48}}
\multiput(88.42,16.35)(0.47,0.17){1}{\line(1,0){0.47}}
\multiput(88.9,16.51)(0.47,0.18){1}{\line(1,0){0.47}}
\multiput(89.37,16.69)(0.23,0.09){2}{\line(1,0){0.23}}
\multiput(89.84,16.87)(0.23,0.1){2}{\line(1,0){0.23}}
\multiput(90.3,17.07)(0.23,0.1){2}{\line(1,0){0.23}}
\multiput(90.76,17.27)(0.23,0.11){2}{\line(1,0){0.23}}
\multiput(91.21,17.49)(0.23,0.11){2}{\line(1,0){0.23}}
\multiput(91.66,17.71)(0.22,0.12){2}{\line(1,0){0.22}}
\multiput(92.11,17.94)(0.22,0.12){2}{\line(1,0){0.22}}
\multiput(92.55,18.18)(0.22,0.12){2}{\line(1,0){0.22}}
\multiput(92.99,18.43)(0.22,0.13){2}{\line(1,0){0.22}}
\multiput(93.42,18.68)(0.21,0.13){2}{\line(1,0){0.21}}
\multiput(93.85,18.95)(0.21,0.14){2}{\line(1,0){0.21}}
\multiput(94.27,19.22)(0.21,0.14){2}{\line(1,0){0.21}}
\multiput(94.69,19.51)(0.21,0.15){2}{\line(1,0){0.21}}
\multiput(95.1,19.8)(0.2,0.15){2}{\line(1,0){0.2}}
\multiput(95.5,20.1)(0.13,0.1){3}{\line(1,0){0.13}}
\multiput(95.9,20.4)(0.13,0.1){3}{\line(1,0){0.13}}
\multiput(96.29,20.72)(0.13,0.11){3}{\line(1,0){0.13}}
\multiput(96.68,21.04)(0.13,0.11){3}{\line(1,0){0.13}}
\multiput(97.06,21.37)(0.12,0.11){3}{\line(1,0){0.12}}
\multiput(97.43,21.71)(0.12,0.11){3}{\line(1,0){0.12}}
\multiput(97.79,22.05)(0.12,0.12){3}{\line(1,0){0.12}}
\multiput(98.15,22.4)(0.12,0.12){3}{\line(0,1){0.12}}
\multiput(98.5,22.76)(0.11,0.12){3}{\line(0,1){0.12}}
\multiput(98.85,23.13)(0.11,0.12){3}{\line(0,1){0.12}}
\multiput(99.19,23.5)(0.11,0.13){3}{\line(0,1){0.13}}
\multiput(99.52,23.88)(0.11,0.13){3}{\line(0,1){0.13}}
\multiput(99.84,24.27)(0.1,0.13){3}{\line(0,1){0.13}}
\multiput(100.15,24.66)(0.1,0.13){3}{\line(0,1){0.13}}
\multiput(100.46,25.06)(0.15,0.2){2}{\line(0,1){0.2}}
\multiput(100.76,25.46)(0.15,0.21){2}{\line(0,1){0.21}}
\multiput(101.05,25.87)(0.14,0.21){2}{\line(0,1){0.21}}
\multiput(101.33,26.29)(0.14,0.21){2}{\line(0,1){0.21}}
\multiput(101.61,26.71)(0.13,0.21){2}{\line(0,1){0.21}}
\multiput(101.87,27.13)(0.13,0.22){2}{\line(0,1){0.22}}
\multiput(102.13,27.57)(0.12,0.22){2}{\line(0,1){0.22}}
\multiput(102.38,28)(0.12,0.22){2}{\line(0,1){0.22}}
\multiput(102.62,28.44)(0.12,0.22){2}{\line(0,1){0.22}}
\multiput(102.85,28.89)(0.11,0.23){2}{\line(0,1){0.23}}
\multiput(103.07,29.34)(0.11,0.23){2}{\line(0,1){0.23}}
\multiput(103.28,29.8)(0.1,0.23){2}{\line(0,1){0.23}}
\multiput(103.49,30.26)(0.1,0.23){2}{\line(0,1){0.23}}
\multiput(103.68,30.72)(0.09,0.23){2}{\line(0,1){0.23}}
\multiput(103.87,31.19)(0.18,0.47){1}{\line(0,1){0.47}}
\multiput(104.04,31.66)(0.17,0.47){1}{\line(0,1){0.47}}
\multiput(104.21,32.13)(0.16,0.48){1}{\line(0,1){0.48}}
\multiput(104.37,32.61)(0.15,0.48){1}{\line(0,1){0.48}}
\multiput(104.51,33.09)(0.14,0.48){1}{\line(0,1){0.48}}
\multiput(104.65,33.57)(0.13,0.49){1}{\line(0,1){0.49}}
\multiput(104.78,34.06)(0.12,0.49){1}{\line(0,1){0.49}}
\multiput(104.9,34.55)(0.11,0.49){1}{\line(0,1){0.49}}
\multiput(105.01,35.04)(0.1,0.49){1}{\line(0,1){0.49}}
\multiput(105.11,35.53)(0.09,0.49){1}{\line(0,1){0.49}}
\multiput(105.2,36.03)(0.08,0.5){1}{\line(0,1){0.5}}
\multiput(105.28,36.52)(0.07,0.5){1}{\line(0,1){0.5}}
\multiput(105.35,37.02)(0.06,0.5){1}{\line(0,1){0.5}}
\multiput(105.41,37.52)(0.05,0.5){1}{\line(0,1){0.5}}
\multiput(105.46,38.02)(0.04,0.5){1}{\line(0,1){0.5}}
\multiput(105.5,38.52)(0.03,0.5){1}{\line(0,1){0.5}}
\multiput(105.53,39.02)(0.02,0.5){1}{\line(0,1){0.5}}
\multiput(105.55,39.52)(0.01,0.5){1}{\line(0,1){0.5}}

\end{picture}

}

\begin{document}

\baselineskip 24pt

\begin{center}

{\Large \bf Gravitational Wave Tails from Soft Theorem:}

{\Large \bf A Short Review}

\end{center}

\vskip .6cm
\medskip

\vspace*{4.0ex}

\baselineskip=18pt

\centerline{\large \rm Ashoke Sen}

\vspace*{4.0ex}

\centerline{\large \it International Centre for Theoretical Sciences - TIFR 
}
\centerline{\large \it  Bengaluru - 560089, India}


\vspace*{1.0ex}
\centerline{\small E-mail:  ashoke.sen@icts.res.in}

\vspace*{5.0ex}

\centerline{\bf Abstract} \bigskip

If a set of massive objects collide in space and the fragments disperse, possibly
forming black holes, then this process
will emit gravitational waves. Computing the detailed gravitational wave-form associated
with this process is a complicated
problem, not only due to the non-linearity of gravity but also due to the fact that during the
collision and subsequent fragmentation the objects could undergo 
complicated non-gravitational
interactions. Nevertheless the classical soft graviton theorem determines the power 
law fall-off of the wave-form at
late and early times, including logarithmic corrections, 
in terms of only the momenta of the incoming and outgoing objects
without any reference to what transpired during the collision. In this short review I shall
explain the results and very briefly outline the derivation of these results.

\vfill \eject

\tableofcontents

\sectiono{Introduction} \label{s0}

Consider a violent explosion in space where a bound system breaks up into many
pieces that fly apart. A detector placed far away will detect gravitational waves emitted
during this explosion after a time that takes the wave to travel from the point of explosion
to the detector. Example of such a process is the explosion of a supernova. We can also
consider a more general situation where a set of objects come together, interact and then
fly apart, possibly exchanging mass, energy, momentum and angular momentum during
this process. This process will also emit gravitational waves. An example of such
a process is provided by the bullet cluster where a pair of galaxy clusters, each of mass
of the order of $10^{14}$ solar mass, pass through each other. 

In general, computation of gravitational waves 
emitted during such processes is complicated.
First of all, when the objects are close they may be subject 
to forces other than the gravitational
force. These could be electromagnetic and or nuclear forces or even 
Fermi pressure.
Second, even when other forces are absent, as in the case
of interaction between black holes, gravitational forces are non-linear and their effect
is usually
studied using numerical relativity or in the weak field approximation.

Surprisingly, some results in quantum theory of gravity, known as the soft graviton theorem,
can be used to get analytical results for some aspects of classical gravitational
radiation emitted during such processes even when the interactions are strong and
interactions other than gravitational interactions are present.
This will be the subject of this review. We shall mainly summarize the results of
\cite{1804.09193,1806.01872,1808.03288,1912.06413,2008.04376}.

\tcr{The particular aspect of gravitational waves that can be studied using soft theorem
is the behaviour of the gravitational wave-form at the detector at late and early retarded
time, i.e.\ long before and long after the bulk of the wave-form produced during
the collision or explosion reaches the detector. These parts of the wave-form are
related to the behaviour of the time Fourier transform of the wave-form at low
frequencies. These in turn are determined from soft theorem which determines the
amplitude for low frequency graviton emission in terms of amplitudes where such
emissions are absent.}  \tcb{The universality of the results follows from the
universality of soft theorem, -- they are consequences of general coordinate invariance
and are independent of what
other interactions are present in the system.}

Throughout this paper we shall work in four space-time dimensions. The analysis
is simpler in larger number of space-time dimensions and can be found in
\cite{1906.08288}. 
Also we shall assume that the incoming and outgoing objects taking part in the
scattering process do not carry any electric charge. If they carry electric charge then
the formul\ae\ given below will have additional terms which can be found in
 \cite{1912.06413,2008.04376}.
 
We shall keep all the fundamental constants in our formul\ae\ except during the analysis
of section \ref{sclassical}. $c$ will denote the speed
of light and $G$ will denote Newton's gravitational constant. We shall also choose the
Minkowski metric
$\eta_{\mu\nu}$ and its inverse $\eta^{\mu\nu}$ to be diag($-1$,1,1,1). All indices will
be raised and lowered by $\eta$. The four momentum of a particle carrying energy $E$
and momentum $\vec p$ will be taken to be $(E/c,\vec p)$ while the space-time
coordinate of the particle will be taken to be $(x^0=c\, t,\vec x)$.

\sectiono{The results} \label{sresults}

Let us consider a set of $m$ charge neutral
objects carrying four momenta $p_1',\cdots, p_m'$ that come together,
interact via gravitational and possibly non-gravitational interactions, and disperse
as a set of $n$ charge neutral objects carrying four momenta $p_1,\cdots, p_n$. 
We denote the masses of these objects by $\{m'_a\}$ and $\{m_a\}$
respectively:
\be
m_a^{\prime2} c^2 = - \eta_{\mu\nu} p_a^{\prime\mu} p_a^{\prime\nu}, \qquad m_a^2 c^2 
= - \eta_{\mu\nu}
p_a^\mu p_a^\nu\, .
\ee
We shall consider a gravitational wave
detector placed at a large distance $R$ away from the region where the interaction takes
place. 
Long before the gravitational wave reaches the detector, 
the metric around the detector is flat up to
corrections of order $R^{-2}$ due to the coulomb field of the incoming objects
and we can choose a coordinate system where the
metric around the detector is $\eta_{\mu\nu}$ up to corrections of order $R^{-2}$.
In this coordinate system, we shall denote the direction
of the scattering region by the unit vector $-\hat n$ so that from the point of
view of the scattering region, the direction of the detector is $\hat n$.
We also introduce the four vector $n$ with components:
\be
n: \quad  (n^0=1,\hat n)\, .
\ee
When the gravitational wave due to the scattering reaches the detector,
the metric around the detector changes by
a term of order $1/R$:
\be
g_{\mu\nu} =\eta_{\mu\nu}+2\, h_{\mu\nu}, \qquad h_{\mu\nu}\sim R^{-1}\, .
\ee
Let $t_0$ be the time when the peak of the signal reaches the detector. The precise
definition of $t_0$ is not important since our formul\ae\ will not be sensitive to shifts
in $t_0$ of order unity. We define
the retarded time $u$ at the detector to be,
\be \label{eut0}
u = t-t_0\, .
\ee
Computing the gravitational wave-form amounts to computing 
the order $R^{-1}$
contribution to $h_{\mu\nu}$ as a function of $u$. This is a complicated problem
and requires detailed knowledge of the scattering process. However the
behaviour of the wave-form at large positive $u$ and large negative $u$ have
concrete expressions in terms of only the four momenta $\{p_a'\}, \{p_a\}$
of the incoming and the
outgoing particles and does not depend on the details of the scattering 
process\cite{1804.09193,1806.01872,1808.03288,1912.06413,2008.04376}..
We shall now present the relevant formul\ae.

We first define 
\be
e_{\mu\nu}\equiv h_{\mu\nu}  - {1\over 2} 
\eta^{\rho\sigma}\, h_{\rho\sigma} \, \eta_{\mu\nu}\, .
\ee
This equation can be reversed as
\be
h_{\mu\nu} = e_{\mu\nu}  - {1\over 2} 
\eta^{\rho\sigma}\, e_{\rho\sigma} \, \eta_{\mu\nu}\, ,
\ee
so that the knowledge of $h_{\mu\nu}$ implies knowledge of $e_{\mu\nu}$ 
and vice versa. The final formul\ae\ are a little easier to write
as expressions for $e_{\mu\nu}$. We have\cite{1808.03288,1912.06413,2008.04376}
(see also \cite{1804.09193,1806.01872} for special cases of these 
formul\ae) 
\be \label{e2.7}
 e_{\mu\nu}= \cases {
 A_{\mu\nu} + B_{\mu\nu} \, u^{-1} + F_{\mu\nu}\,
 u^{-2}\ln u+\OO(u^{-2}) + \OO(R^{-2}), \quad \hbox{for large positive $u$},
 \cr \cr
C_{\mu\nu}\,  u^{-1} + G_{\mu\nu}\,  u^{-2} \ln|u|+
\OO(u^{-2}) +\OO(R^{-2}), \quad \hbox{for large 
negative $u$},
}
\ee
where\footnote{In the $\{(p_a.p_b)^2 
-m_a^2 m_b^2 c^4\}^{3/2} $ factors in the following formul\ae\ we need to choose
the branch with positive sign. If either of the states are massless then this
reduces to $\{(p_a.p_b)^2\}^{3/2} =-(p_a.p_b)^3$ since in our convention
$p_a.p_b<0$.}
\be\label{e2.8}
A^{\mu\nu} = {2\, G\over R\, c^3} \, \left[-\sum_{a=1}^{n}  
p_a^\mu \, p_a^\nu\, 
{1\over n.p_a} + \sum_{a=1}^{m}  
p_a^{\prime\mu} \, p_a^{\prime\nu}\, 
{1\over n.p_a'} \right], \ee
\ben\label{e2.9}
&&{\hskip -.3in  B^{\mu\nu}=-\, {4\, G^2\over R\, c^7} \Bigg[
\sum_{a=1}^n \sum_{b=1\atop b\ne a}^n 
{ p_a.p_b\over 
\{(p_a.p_b)^2 
-m_a^2 m_b^2 c^4\}^{3/2}} \, 
\Bigg\{{3\over 2} m_a^2 m_b^2 c^4- (p_a.p_b)^2\Bigg\} } \non\\ &&
{\hskip 1in 
\times \ {p_a^\mu \over n.p_a}\,
 (n.p_b\, p_a^\nu - n.p_a\, p_b^\nu ) }
\non\\ && {\hskip 1in    \hskip -.5in - 
 \Bigg\{\sum_{b=1}^n\, p_b.n  \sum_{a=1}^{n} \, {1\over p_a.n} \, p_a^\mu p_a^\nu
 - \sum_{b=1}^m\, p'_b.n\sum_{a=1}^{m} \, {1\over p_a'.n} \, p_a^{\prime\mu} p_a^{\prime \nu}\Bigg\}
 \Bigg]\, ,} 
 \een
 \ben\label{e2.10}
 { C^{\mu\nu} }& =& {  {4\, G^2\over R\, c^7} \Bigg[ \sum_{a=1}^m \sum_{b=1\atop b\ne a}^m 
{ p_a'.p_b'\over 
\{(p_a'.p_b')^2 
-m_a^{\prime 2} m_b^{\prime 2}c^4\}^{3/2}} \, 
\left\{{3\over 2} m_a^{\prime 2} m_b^{\prime 2} c^4 - (p_a'.p_b')^2\right\} }\non\\
&& { \hskip .8in \times\ {p_a^{\prime \mu} \over n.p'_a}\,
 (n.p_b^{\prime} \, p_a^{\prime\nu} - n. p_a^{\prime}\, p_b^{\prime\nu} )
 \Bigg]\ \, , }
\een
\ben \label{eaddcon4}
{ F^{\mu\nu}} & {=}&{  2\, {G^3\over R \, c^{11}}\, \Bigg[ 4\, \left\{
\sum_{b=1}^n p_b.n \sum_{d=1}^n p_d.n 
\sum_{a=1}^{n} {p_a^\mu p_a^\nu\over p_a.n} - 
\sum_{b=1}^m p'_b.n \sum_{d=1}^m p'_d.n\sum_{a=1}^{m} {p_a^{\prime\mu} p_a^{\prime\nu}\over 
p_a'.n}\right\} }
\non\\ && \hskip -.7in 
{+ 4\, \sum_{d=1}^n p_d.n \sum_{a=1}^{n} \sum_{b=1\atop b\ne a}^{n}
 {1\over p_a.n} {p_a.p_b\over \{(p_a.p_b)^2 - m_a^2 m_b^2 c^4\}^{3/2}} 
 \{2 (p_a.p_b)^2 - 3 m_a^2 m_b^2 c^4\} \{ n.p_b \, p_a^\mu \, p_a^\nu - n.p_a \, p_a^\mu \, p_b^\nu\}}
 \non\\ &&  \hskip -.9in
{+2\, \sum_{d=1}^m p'_d.n \sum_{a=1}^{m} \sum_{b=1\atop b\ne a}^{m}
 {1\over p'_a.n} {p'_a.p'_b\over \{(p'_a.p'_b)^2 - m_a^{\prime 2} m_b^{\prime 2} c^4\}^{3/2}} 
 \{2 (p_a'.p_b')^2 - 3 m_a^{\prime 2} m_b^{\prime 2} c^4\} \{ n.p'_b \, p_a^{\prime\mu} \, p_a^{\prime \nu} - 
 n.p'_a \, p_a^{\prime \mu} \, p_b^{\prime \nu}\}} \non\\ &&  \hskip -.5in
{+  \sum_{a=1}^{n} \sum_{b=1\atop b\ne a}^{n}
\sum_{d=1\atop d\ne a}^{n}  {1\over p_a.n}  
{p_a.p_b\over \{(p_a.p_b)^2 - m_a^2 m_b^2 c^4\}^{3/2}}   
\{2 (p_a.p_b)^2 - 3 m_a^2 m_b^2 c^4\} {p_a.p_d\over \{(p_a.p_d)^2 - m_a^2 m_d^2 c^4\}^{3/2}} }\non\\ &&
{\{2 (p_a.p_d)^2 - 3 m_a^2 m_d^2 c^4\}  \{ n.p_b \, p_a^\mu  - n.p_a \, p_b^\mu \} \, 
\{ n.p_d \, p_a^\nu - n.p_a \, p_d^\nu \}
\Bigg]\, , }
\een
\ben \label{eaddcon5}
{ G^{\mu\nu}} &{=}& 
{ -2\, {G^3\over R \, c^{11}}\, \Bigg[  2\, \sum_{d=1}^m p'_d.n \sum_{a=1}^{m} \sum_{b=1\atop b\ne a}^{m}
 {1\over p'_a.n} {p'_a.p'_b\over \{(p'_a.p'_b)^2 - m_a^{\prime 2} m_b^{\prime 2} c^4\}^{3/2}} 
 \{2 (p_a'.p_b')^2 - 3 m_a^{\prime 2} m_b^{\prime 2} c^4\} } \non\\ &&  \hskip 1in 
 {\{ n.p'_b \, p_a^{\prime\mu} \, p_a^{\prime \nu} - 
 n.p'_a \, p_a^{\prime \mu} \, p_b^{\prime \nu}\} }\non\\ &&  \hskip -.5in
{-  \sum_{a=1}^{m} \sum_{b=1\atop b\ne a}^{m}
\sum_{d=1\atop d\ne a}^{m}  {1\over p'_a.n}  
{p'_a.p'_b\over \{(p'_a.p'_b)^2 - m_a^{\prime 2} m_b^{\prime 2} c^4\}^{3/2}}   
\{2 (p'_a.p'_b)^2 - 3 m_a^{\prime 2} m_b^{\prime 2} c^4\} {p'_a.p'_d\over \{(p'_a.p'_d)^2 - m_a^{\prime 2} m_d^{\prime 2}
c^4\}^{3/2}}} \non\\ && 
{\{2 (p'_a.p'_d)^2 - 3 m_a^{\prime 2} m_d^{\prime 2} c^4\}  
\{ n.p'_b \, p_a^{\prime\mu} - n.p'_a \, p_b^{\prime\mu} \} \, 
\{ n.p'_d \, p_a^{\prime \nu}  - n.p'_a \, p_d^{\prime\nu} \}
\Bigg]\, .} 
\een
Note that the vanishing of the constant term in the expression for $e_{\mu\nu}$ for
large negative $u$ is a reflection of the coordinate system we have chosen. \tcr{We shall
explain the derivation of  \refb{e2.7}-\refb{eaddcon5}  
in sections \ref{ssoft} and \ref{sclassical}.}

We now make a few remarks on these formul\ae.
\begin{enumerate}
\item We see that the coefficients $A_{\mu\nu}$, $B_{\mu\nu}$, 
$C_{\mu\nu}$, $F_{\mu\nu}$ and $G_{\mu\nu}$ depend only on the momenta of the
incoming and the outgoing particles and not on the details of the scattering process.
\item The coefficient $A_{\mu\nu}$ of the $u$ independent part represents a permanent
change in $h_{\mu\nu}$ that remains even after the gravitational wave has passed. 
\tcb{Therefore a gravitational wave detector does not return to its original
configuration after the passage of the gravitational wave.} This
is the well-known memory effect discovered many years ago\cite{mem1,mem2,mem3,mem4}.
Its relation to soft
theorem\cite{Gell-Mann,low,weinberg1,weinberg2,1404.4091,1706.00759} 
and asymptotic symmetry\cite{Bondi:1962px,Sachs:1962wk}
has been discovered more recently in \cite{1312.2229,1401.7026,1407.3814,
1407.3789,1408.2228,
1411.5745,1502.02318,1505.05346,1506.02906,
1509.01406,1605.09094,1605.09677,1608.00685}.
\item The coefficients $B_{\mu\nu}$ and $F_{\mu\nu}$ describe the late time fall-off
of the gravitational wave-form and the 
coefficients $C_{\mu\nu}$ and $G_{\mu\nu}$ describe the early time growth
of the gravitational wave-form. \tcb{Since these are transient effects that vanish
in the far past and in the far future, }these can be collectively called the tail terms.
They were derived in 
\cite{1804.09193,1806.01872,1808.03288,1912.06413,2008.04376}.\footnote{Such 
terms can also be found by explicit examination of earlier
results in special cases, {\it e.g.} in \cite{peter,Blanchet:1992br}.}

\item Recall that $u$ is defined only up to a shift of order unity since the $t_0$ in 
\refb{eut0} has not been defined precisely.  However one can see that the expansion
in \refb{e2.7} is not affected by this ambiguity, {\it e.g.} under a shift $u\to u+c$, the 
coefficient of the $u^{-1}$ and $u^{-2}\ln u$ terms are not affected. However,
under such a shift the coefficient
of the $u^{-2}$ term will change by a term proportional to $B_{\mu\nu}$ at late
time and by a term proportional to $C_{\mu\nu}$ at early time.
\item 
In the expressions for $ A^{\mu\nu},B^{\mu\nu}$ and $ F^{\mu\nu}$,
the sum over final state particles labelled by $a,b$ includes
integration over the outgoing flux of gravitational 
radiation, regarded as a flux of massless particles. This makes it somewhat
inconvenient to use this formula since unless we can compute the finite frequency
gravitational waves emitted during the scattering, we cannot perform the sum over
final states. 
For $ A^{\mu\nu}$ this had been observed earlier and this effect is known
as  the `non-linear memory'  term\cite{Christodoulou:1991cr,Blanchet:1992br}.
However it turns out that
due to some miraculous cancellations, in $ B^{\mu\nu}$ and $ F^{\mu\nu}$
the contribution from
massless final states can be expressed in terms of massive state momenta.
In particular, in the expression \refb{e2.9} for $ B_{\mu\nu}$, we can
drop massless particles / radiation contribution in the sum over final states, 
and add\cite{1808.03288,2105.08739}
\be \label{enew1}
-\, {4\, G^2\over R\, c^7}  \left[
 P_F^\mu P_F^\nu - P_I^{\mu} P_I^{\nu} \right]
 \ee
 where 
$P_I$ is the total incoming momentum
and $ P_F$ is the total outgoing momentum {\it carried by massive particles}.
Similarly, in the expression \refb{eaddcon4} for $ F_{\mu\nu}$, 
we can drop  massless particles / 
radiation contribution in the sum over final states,
and add\cite{2105.08739}
 \be\label{enew2}
 -{8\, G^3\over R\, c^{11}}\, \left[ 
n.P_F\, P_F^{\mu} \, P_F^{\nu} - n.P_I\, P_I^\mu P_I^\nu\right] \, .
\ee
We note that \refb{enew1}, \refb{enew2} 
are not new formul\ae\ but follow from manipulating the results 
\refb{e2.9}, \refb{eaddcon4}. 
Nevertheless these formul\ae\ are much easier to use since they do not require
knowledge of finite frequency gravitational waves produced during the 
scattering process.
\item We have described the gravitational wave-forms emitted during a scattering
event, but the same formul\ae\ also work in the case of an explosion since this
can be regarded as a special case of scattering when the initial state has just
one object.
In this case \tcr{one can see from \refb{e2.10} and
\refb{eaddcon5} }that $C_{\mu\nu}$ and $ G_{\mu\nu}$ vanish 
and $ e_{\mu\nu}$ takes the form:
\be
e_{\mu\nu} = \cases{A_{\mu\nu} + B_{\mu\nu} \, u^{-1}
+  \, F_{\mu\nu}\,  u^{-2} \ln u\,+ \OO(u^{-2}), \quad \hbox{ for large positive $u$,}\cr\cr
 0, \quad \hbox{ for large 
negative $u$}\, .}
\ee
\item The binary black hole merger can be regarded as a special case of an explosion at
late time. In this case the final state contains just one massive object, -- the final
black hole. We can now use \refb{enew1}, \refb{enew2}, together with \refb{e2.9}
and \refb{eaddcon4} with the final state massless particle contributions removed,
to show that the
net contribution to $B_{\mu\nu}$ and $F_{\mu\nu}$ vanish. Thus during this process
there is no late time $1/u$ and $\ln u/u^2$ tail of the gravitational wave radiation.
This is consistent with the results 
of \cite{PriceI,PriceII,PhysRevD.34.384,9307009,9307010,0309115,
1712.09977,2104.13809,2105.08084,2106.00035,2202.04093,2401.04179,
2401.04170}.
\tcb{In contrast in the merger between a black hole and a neutron star, or between two
neutron stars or in the supernova explosion, some matter is ejected and $B_{\mu\nu}$
and $F_{\mu\nu}$ do not vanish. These may be observable in future gravitational
wave detectors. These may also serve to distinguish a binary black hole merger from
other mergers since in the former process such tail terms are absent. On the other
hand since the memory term is present in all cases, it is most likely to be
discovered first in the process of black hole merger since typically they produce
the strongest signal.
}

\item \tcb{A natural questions is: how big are the tail terms? A quick estimate shows that
if there is a core collapse supernova somewhere 
in our galaxy, then for favorable choice of
parameters the $B_{\mu\nu} /u$ term 
could produce a strain of order $10^{-22}$ at $u$ of 
order one second \cite{1912.06413}. This may be
observable in future gravitational wave detectors.}

\item As already mentioned, we 
have assumed that the incoming and outgoing objects are charge neutral.
If there are charged particles in the initial and final stages there are additional 
contributions to $B_{\mu\nu}$, $C_{\mu\nu}$, $F_{\mu\nu}$ and $G_{\mu\nu}$ which
can be found, {\it e.g.} in \cite{1912.06413,2008.04376}.
\end{enumerate}

It is also useful to express the result \refb{e2.7} in terms of the time Fourier transform
$\tilde e_{\mu\nu}(\omega)$, defined as,
\be\label{efourier}
\tilde e_{\mu\nu}(\omega) = \int du\, e^{i\omega u} \, e_{\mu\nu}(u), \qquad
\tilde h_{\mu\nu}(\omega) = \int du\, e^{i\omega u} \, h_{\mu\nu}(u)\, .
\ee
Then for small $\omega$, $\tilde e_{\mu\nu}$ has an expansion of the
form\cite{1912.06413}:
\be\label{e2.17}
\tilde e_{\mu\nu}(\omega) = i A_{\mu\nu} (\omega+i\eps)^{-1}
- B_{\mu\nu} \ln (\omega + i\eps) + C_{\mu\nu} \ln (\omega-i\eps) 
+{i\over 2} F_{\mu\nu}\, \omega (\ln (\omega+i\eps))^2 
-{i\over 2} G_{\mu\nu} \, \omega (\ln (\omega-i\eps))^2\, ,
\ee
up to terms that are less singular as $\omega\to 0$. There are other ways of
expressing \refb{e2.17} using the relation\cite{1912.06413}
\be\label{e2.18}
2\, \ln(\omega+i\eps) \, \ln(\omega-i\eps)
= (\ln(\omega+i\eps))^2 + (\ln(\omega-i\eps))^2 +
\hbox{less singular terms}\, .
\ee

\sectiono{Outline of the derivation from the soft graviton theorem} \label{ssoft}

The results described in the last section are statements in classical theory of gravity.
Nevertheless, the $u$ dependent parts of the results were first guessed by taking  
classical limits of some results in quantum
theory of gravity, 
known as soft graviton theorem\cite{Gell-Mann,low,weinberg1,weinberg2}.
In this section we shall give a brief outline of the derivation of these
results from soft
graviton theorem\cite{1801.07719,1808.03288,1912.06413,1804.09193,1806.01872}.

In quantum theory of fields the scattering amplitude $A$ 
gives the probability amplitude
for transition of one state to another, with
$|A|^2$ giving transition probability from one state to another.
Let us suppose that
we know the scattering amplitude $A$ for transition from some incoming state $P$ 
to some
outgoing
state $Q$.
The states $P$ and $Q$ could differ in the number of particles they contain, the
momenta and other quantum numbers they carry etc.
Soft graviton theorem gives the transition amplitude for
\be
P \Rightarrow Q + \hbox{some low energy 
gravitons}\, ,
\ee
in terms of the original transition amplitude $A$.

To take the classical limit, we use the following steps:
\begin{enumerate}
\item Take the states $P$ and $Q$ to consist of very massive objects.
\item Compute the transition probability for
\be
 P \Rightarrow Q + \hbox{$M$ low energy gravitons}\, ,
\ee
with each low energy graviton being 
in the energy range $\hbar(\omega,\omega+\delta\omega)$ and
angular range $\delta\Omega$ around $\hat n$.
This takes the form
\be  \label{esoft}
a\, b^M/M!\, ,
\ee
where $a$ and $b$ are constants, with $a$ depending on the details of the scattering
process
but $b$ depends only on initial and final momenta due to
soft graviton theorem. One further finds that in the classical limit $b$ becomes large.
\item \refb{esoft} has a sharp maximum at $M=b$ for large $b$.
This gives the expected classical number of low energy gravitons
in the energy range 
$\hbar(\omega,\omega+\delta\omega)$ and angular range $\delta\Omega$ 
for each $\omega$ and each direction $\hat n$ separately.
\item Once we have found the expected number of  classical soft gravitons this information
can be translated into the time Fourier transform $\tilde h_{\mu\nu}$ of the
gravitational wave-form, leading to
\refb{e2.17}.
\end{enumerate}
There are however some weak points in the analysis that we now list:
\begin{enumerate}
\item In four space-time dimensions the soft graviton theorems, as originally stated,
suffer from infrared divergences beyond the leading order. In order to make sense of the
theorem we have to assume that the infrared cut-off is provided by $\omega$.
This has been verified by explicit computation in quantum 
theory\cite{1808.03288,2308.16807}, 
but this shows that
the soft theorems responsible for \refb{e2.7} are not the original versions of the soft
theorems. 
\item The expected classical number of soft gravitons can be used to determine the 
wave-form only up to a phase. We need separate analysis to determine this phase.
This phase is related to the Coulomb drag on the gravitational wave due to the final
state particles that produces an
extra factor of $e^{i C \omega \ln (\omega R)}$ for some constant $C$ 
in the dependence of $\tilde h_{\mu\nu}$ on the radial variable
$R$. $C$ is calculable from classical analysis and expanding the 
$e^{i C \omega \ln (\omega )}$ factor in a power series in $\omega$ we get
additional terms in the expression for $\tilde h_{\mu\nu}$.
\end{enumerate}
In section \ref{sclassical} we shall describe how \refb{e2.7} can be derived using a direct
classical analysis that does not suffer from these issues.

\sectiono{Direct classical analysis} \label{sclassical}

Our goal in this section will be to find $h_{\mu\nu}$, or equivalently 
$e_{\mu\nu}$,
generated during the collision process, by direct analysis of the classical equations
of motion. For this we follow the approach described in \cite{1912.06413} and
write down the Einstein's equation for the metric in terms of
$e_{\mu\nu}$. To avoid clutter, we shall drop the factors of $c$ in this section.
Without making any approximation, the
equations determining $e_{\mu\nu}$ in the de Donder gauge can be written
as:
\be\label{esimple}
\square\, e^{\mu\nu}(x) = - 
8\, \pi \, G\, T^{\mu\nu}(x)\, ,
\ee
where $\square$ is the standard Laplacian in flat space-time and the right hand side,
by definition, contains all the other terms that would arise from the Einstein's equation.
This will mean that 
$T^{\mu\nu}$ on the right hand side
has explicit dependence on $e^{\rho\sigma}$, and also on 
the trajectories of the various objects
involved in the scattering. It also carries information about all the non-gravitational
interaction terms between matter. 
The trajectories of the objects in turn depend on the metric and hence on 
$e^{\rho\sigma}$.
This leads to a set of complicated non-linear partial differential equations that contains
all the complexities of the Einstein's equations in the presence of matter.

Now, \refb{esimple} can be formally solved as
\be
 e^{\mu\nu}(x) = - 
8\, \pi \, G\, \int d^4 y \, G_r(x, y) \, T^{\mu\nu}(y) \, ,
\ee
where 
$G_r(x,y)$ is the retarded Green's function in flat space-time. We shall choose the
origin of the $y$ coordinate system to be in the region of space-time
where the scattering takes place.
For the Fourier transformed variable defined in \refb{efourier},  one can show using
the known expression for $G_r(x,y)$ that
the above equation takes the following
form for large $|\vec x|$,
\be\label{eintform}
\tilde e^{\mu\nu}(\omega, \vec x) = 
{2 G\over |\vec x|}\, e^{i\omega |\vec x|}\,  \int d^4 y \, e^{-ik.y} \,
T^{\mu\nu}(y), \qquad k=\omega(1, \vec x / |\vec x|)\, .
\ee
We can identify $|\vec x|$ with $R$ and $k$ with $\omega n$ in the notation of
section \ref{sresults}. At this stage this equation is not very useful since the right
hand side has complicated dependence on all the components $\{\tilde
e^{\rho\sigma}\}$ and also on the particle trajectories that have not yet been
determined.

\begin{figure}
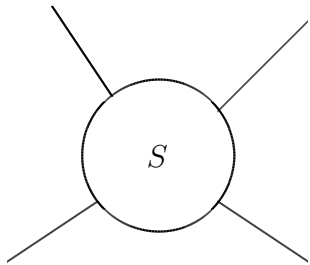

\begin{center}
\figonea
\end{center}
\caption{A schematic representation of a scattering process where the scattering happens
within a region $S$ in space-time and outside this region only long range forces act
on the incoming and outgoing objects.} \label{figone}
\end{figure}

Now we see from \refb{e2.17} that the terms that we want to compute are all non-analytic
for small $\omega$. Here non-analyticity 
does not mean that the term is divergent at 
$\omega=0$, but that if we take sufficient number of derivatives of the term
with respect to $\omega$ and then take the $\omega\to 0$ limit, then the term
diverges. 
Let us now divide the integration region over $y$ in \refb{eintform} into two regions --
the scattering region S  which is a region of large size  around $y=0$ and 
the asymptotic region which is the complement of $S$ (see Fig.~\ref{figone}).
We now note that non-analytic terms in 
$\tilde e^{\mu\nu}$ cannot come from the integral 
of a finite quantity over a finite region. 
Hence the non-analytic terms in $\tilde e^{\mu\nu}$ that we are interested in
are insensitive
to the contribution from the finite region $S$ in $\int d^4 y$.
As a result one can restrict the $y$ integral  in \refb{eintform} to the region outside $S$.
In this region, only long range forces are important.
Furthermore, $e^{\mu\nu}$ is small and as a result $T^{\mu\nu}$ 
simplifies.
The equations can be solved iteratively and give the
result quoted earlier after we supply the missing
factors of $c$\cite{1912.06413,2008.04376}.

For the benefit of the reader we give the set of equations that we need to solve
iteratively, referring the reader to \cite{1912.06413,2008.04376} for the 
details:\footnote{The absence of the familiar $(-\det g)^{-1/2}$ factor in the
expression for $T^{X\mu\nu}$ may seem surprising, but this is due to the fact that while
manipulating Einstein's equation, we take this factor to the left hand side of the
equation\cite{1912.06413,2008.04376}.}
\ben \label{eitone}
&&{ T^{X\mu\nu}(x) \equiv
\sum_{a=1}^n m_a\, \int_0^\infty d\tau\, \delta^{(4)}(x - X_a(\tau)) \, {dX_a^\mu\over d\tau} \, {dX_a^\nu\over d\tau}} \non\\ && \hskip .5in
{+ \ \sum_{a=1}^m
m_a'\, \int_{-\infty}^0 d\tau\, \delta^{(4)}(x - X_a'(\tau)) \, {dX_a^{\prime\mu}\over d\tau} \, {dX_a^{\prime\nu}\over d\tau}\, ,}
\\ \cr
&&{  T^{\mu\nu}(x) = T^{X\mu\nu}(x) 
+ T^{h\mu\nu}(x), } \label{eitoneextra}
\een
\be\label{eittwo}
 \square \, e^{\mu\nu} = -8\, \pi\, G\, T^{\mu\nu}\,  , 
 \ee
 \be \label{eitthree} {
{d^2 X_a^\mu\over d\tau^2} = - \Gamma^\mu_{\nu\rho}(X(\tau)) \,  
{dX_a^\nu\over d\tau} \, {dX_a^\rho\over d\tau} , \qquad  
{d^2 X_a'^\mu\over d\tau^2} = - \Gamma^\mu_{\nu\rho}(X'(\tau)) \,  
{dX_a^{\prime \nu}\over d\tau} \, 
{dX_a^{\prime\rho}\over d\tau}\, .} 
\ee
Here $X_a(\tau)$ and $X'_a(\tau)$ describe the parametric form of the trajectories
of the outgoing and incoming particles respectively, $\tau$ being the 
proper time along the trajectories. $\tau=0$ corresponds to the proper time at
which a given trajectory enters / leaves the region $S$.
The result for massless particles may be obtained
by first computing the result for massive particles and then taking the zero mass
limit. $\Gamma$ denotes the Christoffel symbol constructed from the metric
$\eta_{\mu\nu}+2\, h_{\mu\nu}$. The only equation missing from this list is the expression
for $T^{h\mu\nu}$ in terms of $\{e^{\rho\sigma}\}$. 
This is the gravitational contribution to the stress tensor and may be
found by starting with the Einstein's equation and then taking all the non-linear terms
in $h_{\mu\nu}$ to the right hand side, so that the equation reduces to the form given
in \refb{esimple}. We refer the reader to \cite{1912.06413,2008.04376} for the 
full expression for $T^h$ to the desired order in an expansion in powers of 
$\{e^{\rho\sigma}\}$.

We also need the boundary conditions. For $e_{\mu\nu}$ we use the usual fall-off
condition at past infinity and for the particles trajectories we use,
\be
{ \lim_{\tau\to\infty} 
{dX_a^\mu \over d\tau}=v_a^\mu \equiv {1\over m_a}\, p_a^\mu,}
\qquad {
 \lim_{\tau\to-\infty} 
{dX_a^{\prime\mu} \over d\tau}=v_a^
{\prime \mu} \equiv {1\over m_a'}\, p_a^{\prime\mu}\, .}
\ee
Note that we impose the past boundary conditions on the incoming trajectories
$X_a'(\tau)$ but future boundary condition on the outgoing trajectories
$X_a(\tau)$. This is different from the usual approach to solving classical equations
of motion, where we provide all the boundary conditions at some initial time and
solve the equations of motion for all the variables to find their time evolution. This is
a much more complicated problem, since solving the equations of motion will now
require detailed knowledge of all the forces that act on the particles during their time
evolution. By specifying the mixed boundary conditions we avoid this problem,
\tcr{since we can determine the trajectories of the outgoing particles carrying
momenta $p_1,\cdots,p_n$ by integrating the first set of equations in \refb{eitthree}
backward in time and the trajectories of the incoming particles carrying
momenta $p'_1,\cdots,p'_m$ by integrating the second set of equations in \refb{eitthree}
forward in time. As a result we do not need to integrate the equations in the
scattering region $S$ where non-trivial interactions occur.}

The way we solve the equations \refb{eitone}-\refb{eitthree} is as follows. 
We start with the seed solution where
$e_{\mu\nu}=0$ and the incoming and outgoing trajectories are those of free particles
carrying momenta $\{p_a'\}$ and $\{p_a\}$. We then compute $T^{\mu\nu}$ for this
configuration using \refb{eitone}, \refb{eitoneextra} 
and solve for $e^{\mu\nu}$ using 
\refb{eintform} that follows from \refb{eittwo}. This
already gives the memory term. At the next iteration we correct the trajectories using
\refb{eitthree} and substitute these into \refb{eitone} to correct the expressions for
$T^{X\mu\nu}$. Next we use the known expression for $T^{h\mu\nu}$
to compute $T^{\mu\nu}$ via \refb{eitoneextra}.
We then use this to compute $e^{\mu\nu}$ using \refb{eintform}. This gives the terms
of order $u^{-1}$ in the expression for $e^{\mu\nu}$\cite{1912.06413}. 
Repeating this process
once more we get the expressions for the terms of order $u^{-2}\ln u$\cite{2008.04376}.

\tcb{In the classical analysis,
the expressions for $C_{\mu\nu}$ and $G_{\mu\nu}$ in \refb{e2.10}, \refb{eaddcon5} 
can be
traced to the effect of acceleration of the incoming particles under the influence
of other incoming particles. In the expressions for $B_{\mu\nu}$ and $F_{\mu\nu}$,
the terms in the first two lines of \refb{e2.9} and  last two lines of \refb{eaddcon4}, 
have similar
origin, -- they represent 
the effect of acceleration of the  outgoing particles under the influence
of other outgoing particles. The other terms in the expression for $B_{\mu\nu}$ and
$F_{\mu\nu}$ arise from the backscattering of the memory and the
lower order tail terms in the background gravitational field produced by all the outgoing
particles.
}

\tcb{The universality of the results follow from the fact that we did not need any
information about the interactions in the region $S$ where all the non-universal
terms reside. The ingredients needed for our analysis are the form of the 
trajectory equations given in \refb{eitthree} and the contribution to the gravitational
energy momentum tensor in \refb{eittwo}
from terms with two derivatives of the gravitational field.
Both of these are determined from general coordinate invariance. The only other
effects that can affect the result are other long range forces that do not die off outside
the scattering region. In particular if the incoming and / or outgoing objects are
charged, then they have long range electromagnetic interactions that give
additional contributions to the tail terms.} 

\tcb{The classical analysis described above do not suffer from the
ambiguities that arose in the analysis via soft theorem. The infrared effects are
controlled since while finding iterative solutions to the equations
\refb{eitone}-\refb{eitthree}, we 
have in built infrared cut-off by the positions
of the incoming and outgoing particles and the position of the
detector. The phase ambiguity is absent since we directly compute the
wave-form $h_{\mu\nu}$ instead of the energy flux.}

Given that the classical derivation described above is direct and does not suffer from
having to make the additional assumptions that the derivation based on soft theorem
requires, one might ask whether soft theorems have any role to play in this
analysis. The answer is that although the classical analysis is straightforward, the
question that we answered using the classical analysis is not the kind of question that
we usually ask in the classical theory. There we usually pose a question as an 
initial value problem: given the initial state of the particles, can we predict the final
state and the radiation that is emitted during the scattering? As already discussed,
this is a much more complicated problem. In this sense the soft theorem guides
us towards the kind of questions that have simple answers. Furthermore, although the
analysis based on soft theorem requires some ad hoc assumptions, the procedure
itself is much simpler and gives us a short route to arrive at the final answer. For
example, both for the order $u^{-1}$ term and the order $u^{-2}\ln u$ term the
correct answer was first guessed using the soft theorem and then a more complete
classical derivation was given.

\sectiono{Generalizations} 

The results described above have universal validity independent of the nature of
interactions that govern the scattering as long as gravity is the only long range force.
These results have also been derived by directly analyzing the classical equations
of motion. In this section we shall describe some other results that fail to satisfy one
of the criteria mentioned above but nevertheless are important.

\subsection{Spin dependence}

\def\Si{S}

The coefficients $A^{\mu\nu}$, $B^{\mu\nu}$, $C^{\mu\nu}$, $F^{\mu\nu}$ and
$G^{\mu\nu}$ are independent of the intrinsic angular momenta carried by
the incoming and the outgoing objects. Based on soft theorem, one can also
predict the spin dependence of the gravitational wave-form 
and show that it
should appear in the coefficient of the order $u^{-2}$ 
term  in $e^{\mu\nu}$ or equivalently $\OO(\omega\ln\omega)$
term in $\tilde e^{\mu\nu}$.
The correction to $\tilde e^{\mu\nu}$ takes the form\cite{2106.10741}: 
\ben\label{espin1}
\Delta e^{\mu\nu} 
&=& {G^2\over R\, c^8} \, i\omega\ln(\omega+i\eps)
\, \Bigg[2 \sum_{b=1}^n p_b.n\, 
\Bigg\{\sum_{a=1}^n {p_a^\mu n_\rho \over p_a.n}\Si_a^{\rho\nu}
+ {p_a^\nu n_\rho \over p_a.n}\Si_a^{\rho\mu} \Bigg\} \nonumber \\ && 
- 2 \sum_{b=1}^m p'_b.n\,
\Bigg\{\sum_{a=1}^m {p_a^{\prime\mu} n_\rho \over p'_a.n}\Si_a^{\prime\rho\nu}
+ \sum_{a=1}^m {p_a^{\prime\nu} n_\rho \over p'_a.n}\Si_a^{\prime\rho\mu}
\Bigg\} \nonumber \\
&+& \sum_{a=1}^n \sum_{b=1\atop b\ne a}^n 
{ p_a.p_b (2 (p_a.p_b)^2 -3 m_a^2 m_b^2 c^4)\over 
\{(p_a.p_b)^2 
-m_a^2 m_b^2 c^4\}^{3/2}} {n_\rho n_\sigma\over p_a.n}\, 
\{(p_a^\mu p_b^\rho - p_a^\rho p_b^\mu) \Si_a^{\sigma\nu}
+ (p_a^\nu p_b^\rho - p_a^\rho p_b^\nu) \Si_a^{\sigma\mu}\}\Bigg]\nonumber \\
& +& {G^2\over R\, c^8} \, i\omega \ln(\omega-i\eps) \, 
\sum_{a=1}^m \sum_{b=1\atop b\ne a}^m 
{ p'_a.p'_b (2 (p'_a.p'_b)^2 -3 m_a^{\prime 2} m_b^{\prime 2} c^4)\over 
\{(p'_a.p'_b)^2 
-m_a^{\prime 2} m_b^{\prime 2} c^4\}^{3/2}} {n_\rho n_\sigma\over p'_a.n}
\nonumber \\ && \hskip .5in \times
\{(p_a^{\prime\mu} p_b^{\prime\rho} - p_a^{\prime\rho} 
p_b^{\prime\mu}) \Si_a^{\prime\sigma\nu}
+ (p_a^{\prime\nu} p_b^{\prime\rho} - p_a^{\prime\rho} p_b^{\prime\nu}) \Si_a^{\prime\sigma\mu})\}, 
\een
where $\Si_a^{\prime\mu\nu}$ and $\Si_a^{ \mu\nu}$ are the intrinsic spins
carried by the incoming and outgoing particles respectively, written in the 
relativistic notation.
It is also possible to drop the sum over massless final states inside the
first square bracket in \refb{espin1}, 
by adding the contribution\cite{2106.10741}
\be
-2 {G^2\over R\, c^8} \, i\, \omega \ln (\omega+i\eps)
\, \left[ P_F^\mu \, n_\rho \, {\sum_a}' \Si_a^{\rho\nu}
+ P_F^\nu \, n_\rho \, {\sum_a}' \Si_a^{\rho\mu} - 
P_I^\mu \, n_\rho \, \sum_{a=1}^m \Si_a^{\prime\rho\nu}
- P_I^\nu \, n_\rho \, \sum_{a=1}^m \Si_a^{\prime \rho\mu} 
\right]\, ,
\ee
where $\sum'$ denotes sum over all massive final state particles and, as before,
$P_I$ denotes the total momentum carried by all the incoming particles and
$P_F$ denotes the total momentum carried by all the outgoing {\it massive}
particles. 

From \refb{espin1} we can derive the late and early time behaviour of $e^{\mu\nu}$
by noting that the coefficient of the $i\omega\ln(\omega+i\eps)$ 
term in $\tilde e^{\mu\nu}$
gives the coefficient of the $-1/u^2$ term in $e^{\mu\nu}$ for large positive $u$ and
the coefficient of the $i\omega\ln(\omega-i\eps)$ term in $\tilde e^{\mu\nu}$
gives the coefficient of the $1/u^2$ term in $e^{\mu\nu}$ for large negative $u$.
These are subdominant compared to 
the terms that have been written down explicitly
in \refb{e2.7}.
However there are also spin independent contributions at this order and they
suffer from ambiguities that can 
only be resolved by knowing the details of the scattering process. 
Also, we see that a shift of $u$ by a constant in the term
proportional to $1/u$ in \refb{e2.7} will generate
a term of order $u^{-2}$.
Hence the meaning
of spin dependent contribution becomes unclear in general.

However, it was shown in \cite{2106.10741} 
that if we consider the case where the interaction
between the objects is purely gravitational and expand the result in powers of $G$
at fixed values of the impact parameters,
then the ambiguity in the $u^{-2}$ term is of order $G^3$ whereas the
spin dependent contribution predicted by soft graviton theorem is of order $G^2$.
Hence in this case
we can in principle test the prediction for the spin dependent term at order
$G^2/u^2$.
This part of the spin dependent term was
reproduced by direct classical analysis in \cite{2106.10741}. \tcb{As in the case
of the tail terms proportional to $u^{-1}$ and $u^{-2}\ln u$, the spin dependent tail terms
proportional to $u^{-2}$ also vanish for binary black hole merger when there is
only one massive object in the final state \cite{2106.10741}.}

\subsection{An all order prediction}

Given the universality of the results \refb{e2.7}-\refb{eaddcon5}, one can ask if there are
generalizations of these results to higher order. In particular we can consider the 
series of terms proportional to $u^{-k}(\ln u)^{k-1}$ in the expression
for $e^{\mu\nu}$ for large $|u|$ or equivalently 
$\omega^{k-1} (\ln\omega)^k$ terms in the expression for
$\tilde e^{\mu\nu}$ for small $\omega$. We can begin with a general
postulate that $\tilde e^{\mu\nu}$ is given by the small $\omega$ expansion 
of\cite{2008.04376}:\footnote{To 
determine the $u$ dependence of $e^{\mu\nu}$ from these
equations, we first
make repeated use of \refb{e2.18} to arrive at the relation
$
(\ln(\omega+i\eps))^m (\ln(\omega-i\eps))^n
= {m\over m+n} (\ln(\omega+i\eps))^{m+n}+ {n\over m+n} (\ln(\omega-i\eps))^{m+n}
+\hbox{less singular terms}
$,
and then use this relation to express the small $\omega$ 
expansion of $\tilde e^{\mu\nu}$ as linear combinations of $\omega^{k-1}
(\ln(\omega+i\eps))^k$ and $\omega^{k-1}
(\ln(\omega-i\eps))^k$. After taking the Fourier transform, the coefficient of the
$\omega^{k-1}
(\ln(\omega+i\eps))^k$ term in $\tilde e^{\mu\nu}$
determines the coefficient of the $u^{-k}(\ln u)^{k-1}$ term in $e^{\mu\nu}$ for
large positive $u$ and the coefficient of the
$\omega^{k-1}
(\ln(\omega-i\eps))^k$ term  in $\tilde e^{\mu\nu}$ 
determines the coefficient of the $u^{-k}(\ln |u|)^{k-1}$ term in $e^{\mu\nu}$ for
large negative $u$.}
\ben\label{e5.4}
\tilde e^{\mu\nu} &=& - 
\exp\left\{-{2\, i\, G c^{-4}} \,\omega\, \ln\, (\omega+i\eps)\,
 {\sum_a}p_a.n \right\}\,  \sum_{k\ge 0}{1\over k!}
  A^{\mu\nu}_{(k)} \, (-i\omega)^{k-1} (\ln\,
(\omega+i\eps))^k
\nonumber \\ &&
 - 
\exp\left\{-{2\, i\, G c^{-4}} \,\omega\, \ln\, (\omega+i\eps)\,
 {\sum_a}p'_a.n \right\}\,  \sum_{k\ge 0}{1\over k!}
  B^{\mu\nu}_{(k)} \, (i\omega)^{k-1} (\ln\,
(\omega-i\eps))^k\, ,
\een
where
\be \label{eamunudef}
A^{\mu\nu}_{(0)} ={2\, G\over R\, c^3} \,{\sum_{a}}
p_a^\mu \, p_a^\nu\, 
{1\over n.p_a}, \qquad B^{\mu\nu}_{(0)} ={2\, G\over R\, c^3} \, \sum_{a}
p_a^{\prime\mu} \, p_a^{\prime\nu}\, 
{1\over n.p_a'} \, ,
\ee
\be \label{ememag}
A^{\mu\nu}_{(1)} =  {(2\, G)^{2}\over  R\, c^{7}} \, {\sum_{a,b}}
{\tau_{ab}
  \over p_a.n} n_\rho \, p^\mu_{[a} p^\rho_{b]}\, 
 p^\nu_{a} \, ,
\qquad
B^{\mu\nu}_{(1)} = {(2\, G)^{2}\over  R\, c^{7}} \, \sum_{a,b} 
{\tau'_{ab}
  \over p'_a.n} n_\rho \, p^{\prime\mu}_{[a} p^{\prime\rho}_{b]}\, 
 p^{\prime\nu}_{a} \, ,
\ee
\ben
\tau_{ab} &=& {1\over 2}\, p_a.p_b\ {2(p_a.p_b)^2 - 3 m_a^2 m_b^2 c^4\over
\{(p_a.p_b)^2 - m_a^2 
m_b^2 c^4\}^{3/2}}\, ,
\nonumber \\
\tau'_{ab} &=& {1\over 2}\,  p'_a.p'_b\ {2(p'_a.p'_b)^2 - {3} m_a^{\prime 2} m_b^{\prime 2} 
c^4\over  \{(p'_a.p'_b)^2 - m_a^{\prime 2} 
m_b^{\prime 2} c^4\}^{3/2}}  \, ,
\een
\be
P_{[a}Q_{b]} \equiv  (P_aQ_b-P_bQ_a)\, ,
\ee
\be\label{eguessa2}
A^{\mu\nu}_{(2)} = {(2\, G)^{3}\over  R\, c^{11}}
{\sum_{a,a_1,a_2} }
{\tau_{aa_1}
 \tau_{aa_2} \over p_a.n} n_\rho \, p^\mu_{[a} p^\rho_{a_1]}\, 
n_\sigma p^\nu_{[a} p^\sigma_{a_2]} 
n_{\alpha_3}\, ,
\ee
\be\label{eguesstwob2}
B^{\mu\nu}_{(2)} ={(2\, G)^{3}\over R\, c^{11}} 
\sum_{a,a_1,a_2} {\tau^{\prime}_{aa_1}
 \tau^{\prime}_{aa_2}\over p_a.n} n_\rho \, p^{\prime\mu}_{[a} p^{\prime\rho}_{a_1]}
\,
n_\sigma \, p^{\prime\nu}_{[a} p^{\prime\sigma}_{a_2]}
\, ,
\ee
and $A^{\mu\nu}_{(k)}$ and $B^{\mu\nu}_{(k)}$ for $k\ge 3$
are yet to be determined. 
By expanding the exponential term in \refb{e5.4} 
and using \refb{e2.18}, one can verify
that \refb{e5.4} reproduces the coefficients of the $1/\omega$,
$\ln\omega$ and $\omega (\ln\omega)^2$ terms in \refb{e2.17}
with the values of the coefficients $A_{\mu\nu}$,
$B_{\mu\nu}$, $C_{\mu\nu}$, $F_{\mu\nu}$ and $G_{\mu\nu}$ given in 
\refb{e2.8}-\refb{eaddcon5}.
Indeed,
the constants $A^{\mu\nu}_{(k)}$ and  
$B^{\mu\nu}_{(k)}$  for $0\le k\le 2$  have been adjusted to reproduce 
\refb{e2.17}. 
For the $\omega^{-1}$ term
in the expansion of \refb{e5.4} in powers of $\omega$ we take the $i\eps$ prescription
to be $(\omega+i\eps)^{-1}$, -- otherwise the $i\eps$ prescription in the $\omega^k$
term is not important.
The overall phase in \refb{e5.4} could have been
expanded in powers of $\omega\ln\omega$ and absorbed into the definition of
$A^{\mu\nu}_{(k)}$ and $B^{\mu\nu}_{(k)}$, but it is more natural to 
separate it out\cite{2008.04376} since it
comes from the effect of the Coulomb drag on the outgoing gravitational waves.
In the absence of concrete proposals for $A^{\mu\nu}_{(k)}$ and $B^{\mu\nu}_{(k)}$
for $k\ge 3$, \refb{e5.4} contains the same information as \refb{e2.7}-\refb{eaddcon5}.

Ref.~\cite{2407.04128} 
proposed specific forms of $A^{\mu\nu}_{(k)}$ and $B^{\mu\nu}_{(k)}$
for $k\ge 3$. However, before we describe this,  let us make a few observations:
\begin{enumerate}
\item Gauge invariance requires $ \tilde e^{\mu\nu}$ to satisfy the conditions
$n_\mu \, \tilde e^{\mu\nu}=0$. This can be achieved by demanding that 
\be\label{egaugecon}
n_\mu (A^{\mu\nu}_{(0)} - B^{\mu\nu}_{(0)})=0, \qquad
n_\mu \, A^{\mu\nu}_{(k)}=0, \quad n_\mu \, B^{\mu\nu}_{(k)}=0 
\quad \hbox{for $k\ge 1$}\, .
\ee
\refb{eamunudef}-\refb{eguesstwob2} automatically satisfy these constraints, but
any proposal for $k\ge 3$ must obey these conditions, Furthermore 
$A^{\mu\nu}_{(k)}$ and $B^{\mu\nu}_{(k)}$ must also be symmetric under the 
exchange of $\mu,\nu$.
\item In the expression for $\tilde e^{\mu\nu}$, each factor of $\omega$ (except
those appearing in the argument of the logarithm) is accompanied by
$n$. Therefore in the expressions for
$A^{\mu\nu}_{(k)}$ and $B^{\mu\nu}_{(k)}$, 
the number of factors of $n$ appearing in the numerator  minus similar 
factors appearing in the denominator must be equal to $k-1$.
\item Even though there are higher order soft theorems
\cite{1801.05528,1802.03148}, they do not determine the form of
$A^{\mu\nu}_{(k)}$ and $B^{\mu\nu}_{(k)}$ for $k\ge 3$
even if we follow the approach outlined in section \ref{ssoft}. Instead 
they constrain only the projection of $\tilde e^{\mu\nu}$ along
certain subspace which essentially amounts to imposing \refb{egaugecon}. 
\item For large $u$, terms of order $u^{-k-1}(\ln u)^{k}$ are subdominant compared to
the terms we have ignored. For example $u^{-3} (\ln u)^2$ will be subdominant compared
to $u^{-2}$. Therefore one may wonder if there is any significance of the $u^{-k-1}
(\ln u)^k$ terms that we are analyzing. To this end note that  by applying appropriate
differential operator involving $\p/\p u$ on 
$e^{\mu\nu}$ we can eliminate the apparently more dominant terms and
ensure that $u^{-k-1}
(\ln u)^k$ term gives the dominant contribution in the large $u$ limit. For example 
$(u \, e^{\mu\nu})'$, where $'$ denotes derivative with respect to $u$, eliminates the
$u^{-1}$ term and 
$(u(u^2 (u \, e^{\mu\nu})')')'$ eliminates the $u^{-1}$, $u^{-2}$ and $u^{-2}\ln u$ terms in the
expression for $e^{\mu\nu}$. Therefore $(u(u^2 (u \, e^{\mu\nu})')')'$ 
receives the dominant 
contribution from the $u^{-3} (\ln u)^2$ term in the expression for $e^{\mu\nu}$. 
This way, by applying higher derivative operators
on $e_{\mu\nu}$, we can construct expressions for which $u^{-k-1}
(\ln u)^k$ term gives the dominant contribution in the large $u$ limit. 
\end{enumerate}

Let us now describe the proposed 
forms of $A^{\mu\nu}_{(k)}$ and $B^{\mu\nu}_{(k)}$
for $k\ge 3$.
Drawing inspiration from  an earlier conjecture
in \cite{2008.04376}, the gauge invariance condition \refb{egaugecon} and
the analysis of purely gravitational scattering of massive objects, ref.~\cite{2407.04128}
conjectured a general from for the functions $A^{\mu\nu}_{(k)}$ and $B^{\mu\nu}_{(k)}$
for $k\ge 3$.
They take the form:\footnote{The exact statement of the conjectures given here
differs somewhat from \cite{2407.04128}. For example the $i\eps$ prescriptions were
not given in  \cite{2407.04128}. Also \refb{eext} is slightly
more general than what has been 
given in \cite{2407.04128}. Previous attempts to extend the
result beyond the subsubleading order in special cases
has been made in \cite{1903.12419}.}
\be\label{eguess}
A^{\mu\nu}_{(k)} = { (2\, G)^{k+1}\over  R\, c^{3+4k}}
{\sum_{a,a_1,a_2,\cdots, a_k\atop a_1,\cdots, a_k\ne a} }
{\tau_{aa_1}
 \tau_{aa_2} \cdots
\tau_{aa_k} \over p_a.n} n_\rho \, p^\mu_{[a} p^\rho_{a_1]}\, 
n_\sigma p^\nu_{[a} p^\sigma_{a_2]} 
n_{\alpha_3} (p^{\alpha_3}_a+p^{\alpha_3}_{a_3}) \cdots
n_{\alpha_k} (p^{\alpha_k}_a+p^{\alpha_k}_{a_k})\, ,
\ee
\be\label{eguesstwo}
B^{\mu\nu}_{(k)} ={(2\, G)^{k+1}\over R\, c^{3+4k}} 
\sum_{a,a_1,a_2,\cdots, a_k\atop a_1,\cdots, a_k\ne a} {\tau^{\prime}_{aa_1}
 \tau^{\prime}_{aa_2} \cdots
\tau^{\prime}_{aa_k} \over p'_a.n} n_\rho \, p^{\prime\mu}_{[a} p^{\prime\rho}_{a_1]}
\,
n_\sigma \, p^{\prime\nu}_{[a} p^{\prime\sigma}_{a_2]}
n_{\alpha_3} (p^{\prime\alpha_3}_a+p^{\prime\alpha_3}_{a_3}) \cdots
n_{\alpha_k} (p^{\prime\alpha_k}_a+p^{\prime\alpha_k}_{a_k})\, .
\ee
It has also been conjectured that except
for the coefficient of the $\omega^{-1}$ term in the expansion, \refb{e5.4}
admits a rewriting where in the sum over the final state particles we  drop the
massless state contributions and add to $A^{(0)}_{\mu\nu}$ and  
$B^{(0)}_{\mu\nu}$ the terms
\be\label{eext}
\delta A_{(0)}^{\mu\nu} =  -{2 \, G\over R\, c^3} {P_F^\mu\, P_F^\nu\over n.P_F}\, , \qquad
\delta B_{(0)}^{\mu\nu} = - {2 \,  G\over R\, c^3} {P_I^\mu\, P_I^\nu\over n.P_I}\, , 
\ee
where, as before, $P_I$ is the total incoming momentum and $P_F$ is the total
momentum carried by the massive outgoing states.

Evidence for \refb{eguess}, \refb{eguesstwo}  was provided in 
\cite{2407.04128} by working at the leading order in the post Newtonian
limit for $2\to 2$ scattering.  Other than this, there have been no tests or derivations
of these conjectures. Neither has the compatibility of \refb{eguess}, \refb{eguesstwo}
and \refb{eext} been checked.
Preliminary investigation with Biswajit Sahoo suggests that these relations are internally
consistent for $2\to 2$ scattering, but the situation is more complicated 
when the number of incoming or outgoing particles exceeds two.
It will be interesting to reach a definite conclusion.

\sectiono{Related developments}

We end this review with a discussion of some other related developments.

\begin{enumerate}

\item
The results described in this review give us the  low
frequency behaviour of the Fourier transform of the wave-form. This has been
used to compute the low frequency behaviour of the flux of gravitational wave radiation
during a scattering process\cite{2105.08739,2407.04128},  qualitatively
confirming an earlier
conjecture of \cite{1409.4555,1812.08137,1901.10986} on the $\omega$ 
dependence of the flux.
\item
While calculating the full gravitational wave-form produced in a general scattering 
process
is hard, in special cases, {\it e.g.} when the only interactions are gravitational and the
gravitational field remains weak so that the non-linearities can be treated perturbatively,
one can compute the gravitational wave form without making the low frequency 
approximation. Nevertheless, if we take a low frequency limit of these results, it must
agree with the results obtained using soft theorem. This has been tested in many
examples\cite{1901.10986,2007.02077,2211.13120,
2309.14925,2312.07452,2402.06361,2402.06533,2402.06604,
2407.02076,2407.04128,2408.07329}. \tcr{In these papers the authors use different
analytical methods to compute the Fourier transform of the 
gravitational waveform emitted during a scattering process, not
necessarily in the soft limit. One then studies the behaviour of these wave-forms
in the low frequency limit and compares with the formulae\ given in this review,
finding perfect agreement.} 

\tcb{In a similar spirit, the tail terms could also serve as a check
on the results of numerical relativity. Since for binary black hole merger the tail
terms are absent, we do not get any extra information from these terms.
But for the merger of two
neutron stars or a black hole and a neutron star the tail terms are present and one
could use them as a check on the numerical results once they are sensitive enough to
capture the tail terms.}

\item
As already mentioned, the memory effect is related to the BMS 
super translation symmetry and the leading soft graviton theorem.
For tree level scattering amplitudes this relation was extended to subleading order
in \cite{1407.3814,1407.3789,1506.02906,1505.05346,1605.09677}. However, 
as discussed above, once we
take into account the loop contribution to the scattering amplitudes, the subleading and
higher order terms begin to  involve logarithms of the graviton energy and
the usual versions of the soft theorem break down. So the question is whether the 
logarithmic terms in the soft theorem can be accounted for in a modified version of the
asymptotic symmetries. This has been explored  in 
\cite{1903.09133,1912.10229,2205.11477,2309.11220,2403.13053,2407.07978} for the
subleading soft theorem. \tcb{This subject is still being developed, and could
eventually lead to a deeper understanding of asymptotic symmetries.}
\end{enumerate}

\medskip

\noindent {\bf Acknowledgement:} I wish to thank Alok Laddha, Arnab Priya Saha and
Biswajit Sahoo for collaborations and Alok Laddha and Biswajit Sahoo for discussions and
comments on an earlier version of the manuscript.
This work was supported by the ICTS-Infosys Madhava 
Chair Professorship
and the Department of Atomic Energy, Government of India, under project no. RTI4001.

\end{document}